\documentclass[10pt,twoside]{article}
\usepackage{graphicx}
\def\a4{\hsize 17.0cm \vsize 25.cm}

\makeindex

\begin{document}

\title{Gamow Legacy and the Primordial Abundance of Light Elements}

\markboth{E.\,Terlevich}{Primordial Abundance of Light Elements}

\author{Elena Terlevich$^{1}$, Roberto Terlevich$^{1}$ and Valentina Luridiana$^{2}$
\\[5mm]
\it $^{1}$Instituto Nacional de Astrof\'\i sica Optica y Electr\'onica,  Puebla, 
M\'exico;\\
\it e-mail: eterlevi@inaoep.mx, rjt@inaoep.mx \\
\it $^{2}$ Instituto de Astrof\'\i sica de Andaluc\'\i a, Granada, Spain;\\
\it e-mail: vale@iaa.es
}

\date{}
\maketitle

\thispagestyle{empty}

\begin{abstract}
\noindent  The presently accepted ``Theory of the Universe"
was pioneered 60 years ago by Gamow, Alpher and Herman. As a consequence
of the, later dubbed, Hot Big-Bang, matter was neutrons, and after some
decay protons, and a history of successive captures built up
the elements.

It wasn't until some 15 years later (with the discovery of the
Cosmic Microwave Background radiation) that Gamow and colleagues theories
were validated and present day Standard Big-bang Nucleosynthesis theory
was developed.

We will discuss the importance of state of the art observations and
modelling in the quest to determine precise values of the primordial
abundance of D and $^4$He, using observations of astrophysical
objects and modern day atomic parameters. In particular, we will present
the search for understanding and coping with systematic errors
in such determinations.

\end{abstract}

\section{Introduction}

The principles of the Standard Big Bang (SBB) model of the universe were
laid down by Gamow, Alpher and Herman in work produced between 1945 and 1953
(\cite{1, 2, 3, 4}).  Their work 
placed Friedmann's and Lemaitre's conception of the universe on a  physically 
observable basis. They had the advantage of knowing already about the expansion of the 
universe, discovered by Hubble (\cite{5}). Their ideas proposed a way to create 
all the chemical elements.

SBB states that the Big Bang was hot and dense, and
that the conditions were ripe for nucleosynthesis during a short
time window of a few minutes. Before that moment the radiation was
too hard to permit any nucleide to survive, after that the temperature
had gone too low. Therefore the first light nucleides were formed
only in a time window of approximately 15 minutes, and the relative 
abundances that result depend on the production and the destruction
rates during the nucleosynthesis period, which are a function of the
baryon-to-photon ratio $\eta$ (see, e.g.~\cite{6}).

Hayashi in 1950 (\cite{7}) 
suggests that at the high density and temperature reached
close to the discontinuity, a freezing of the ratio neutrons to protons would have 
occurred. Considering as well that there are no stable nuclei with 
atomic weight A=5,8, lead researchers to the conclusion that there was
 no nucleosynthesis after  $^4$He. Successes in stellar evolution and 
nucleosynthesis then lead to Big Bang Nucleosynthesis theory (BBNS) to be 
all but  abandoned for 10 years, until Penzias and Wilson (\cite{8}) 
discovered Cosmic Microwave Background radiation (CMB) that had already been
predicted  by Gamow and colleagues. People then turned back their attention to
Standard Big Bang Nucleosynthesis (SBBN) and a very comprehensive model was developed by
Peebles, Wagoner, Fowler, Hoyle, Schramm, Steigman, etc. (e.g.~\cite{6}). 

A schematic illustration of how the first  light nucleides were synthesized 
in a very short time interval following the Big Bang is shown in Figure 1.

\begin{figure}[h]
\centering
\includegraphics
[height=9cm,width=12.cm]{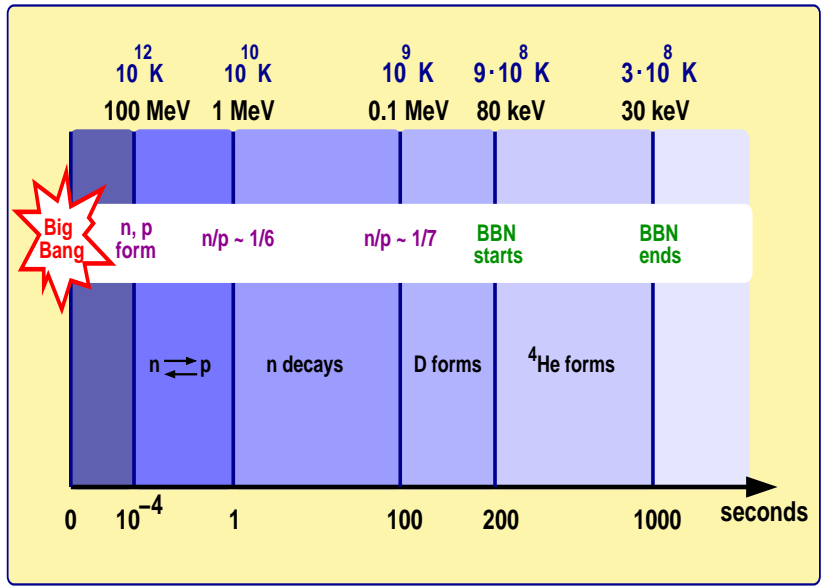}
\caption[h]{The figure shows how the first light nucleids where synthesized in the 
first 1000 seconds after the Big Bang. Deuterium, $^4$He and traces of $^7$Li and Bo
nuclei were formed. After 1000 seconds, and when temperature was 3.10$^8$ degrees, BBN 
stopped.}
\label{fig1}
\end{figure}

SBBN has been very successful at least in three predictions:

\begin{itemize}
\item it very well predicts primordial abundances of
Deuterium, Helium 4, Helium 3, Lithium 7 covering
9 orders of magnitude, a commendable achievement.
\item it predicts the number of neutrino species.
\item it predicts the neutron half life.
\end{itemize}

There is a very delicate balance between the reaction rates and the expansion of the 
universe, therefore on the density and temperature, and hence, the primordial 
abundances of the
first elements can be used to determine the baryon-to-photon 
density $\eta$.  Figure 2 shows the relationship between the primordial abundances of 
$^4$He/H,  $^3$He/H, D/H
and $^7$Li/H and the ratio of the density of baryons-to-photons in units of 10$^{-10}$ 
($\eta_{10}$). One element alone should be enough to determine $\eta$, but consistency
with the others would make the theory very robust, while confirming the parameters
of atomic physics used in  deriving  the predictions.

Of all the light elements, $^4$He is the easiest to measure, but, as Figure 2 shows, 
it is the least sensitive to $\eta$, and therefore, in order to be used as a 
diagnostic, it has to be measured with very high precision.

\begin{figure}[h]
\centering
\includegraphics
[height=9cm,width=9.5cm]{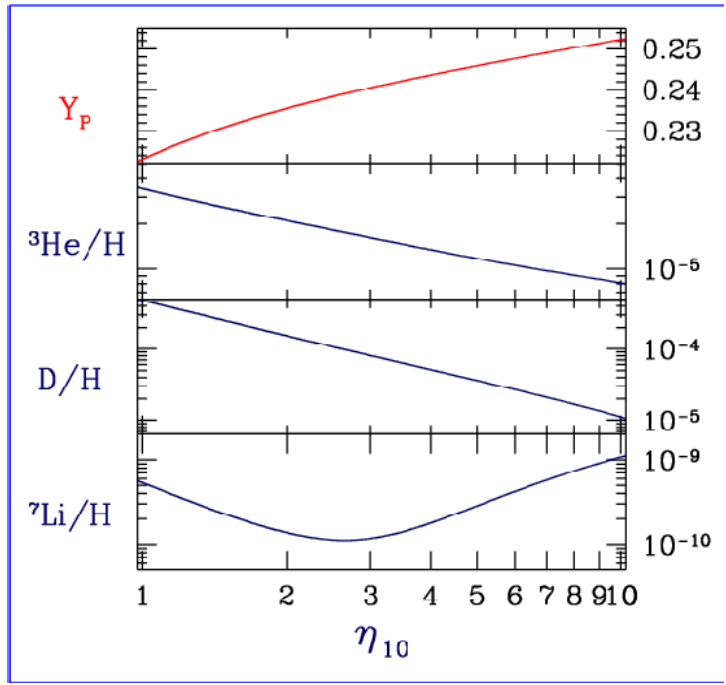}
\caption[h]{Relationship between the primordial abundances of $^4$He/H,  $^3$He/H, D/H
and $^7$Li/H and the ratio of the density of baryons-to-photons in units of 10$^{-10}$ 
($\eta_{10}$). Notice on the right, the 9 orders of magnitude that the abundances 
cover.}
\label{fig2}
\end{figure}

\section{Primordial helium (Y$_p$)}

Abundances of $^4$He and its primordial value (Y$_p$), have been  inferred 
for many years by different authors. Analysing their results, it becomes
apparent that the different determinations are progressively 
converging, although a significant scatter remains. In order to see 
where this scatter comes from, we will review the method used to determine 
Y$_p$.

\subsection{The method}

\begin{figure}[h]
\centering
\includegraphics
[height=9cm,width=9.5cm]{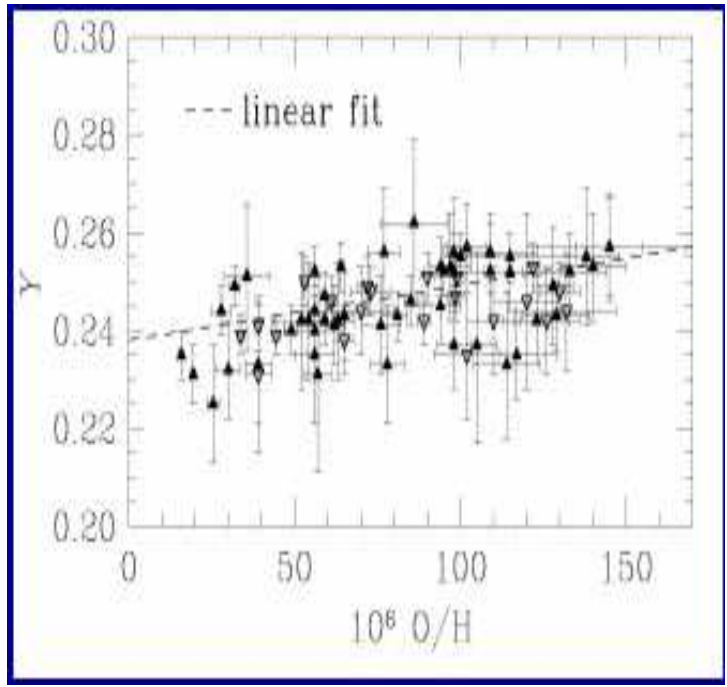}
\caption[h]{Helium abundance (Y) versus oxygen abundance (O/H) for HII galaxies. 
The dashed line is a linear fit.
Figure taken from Fields \& Olive \cite{9}.}
\label{fig3}
\end{figure}

The method, essentially devised by Manuel Peimbert and Silvia Torres-Peimbert in 1974
(\cite{10})  was based on the realization, by Searle and Sargent (\cite{11})
 that
the so called HII galaxy most metal poor known (then and still now, more than 30 years 
later, but that is another story) -- IZw~18 -- still has a finite He/H abundance, 
around 0.24, which therefore had to be primordial.
The thought behind \cite{10,11} is that if the universe was born
with zero metallicity (Z=0), the metals we see today were created by nucleosynthesis 
inside stars, and by the same process, helium abundance grows. 
Therefore,
if we collect a sample of objects, measure their chemical abundances, and 
extrapolate the relation Y(Z) back to Z=0, we can determine Y$_p$.
An up-to-date representation of this concept can be seen in figure \ref{fig3} from a 
compilation by  \cite{9}.
It is clear from the figure that a reliable determination of Y$_p$ rests 
entirely on accurate values of Y and O/H abundances.

\subsection{ $^4$He abundance}

Many efforts have concentrated over the years on Y$_p$ determinations, and although 
the value obtained seems to be converging, substantial scatter still remains.
Given that the determination of  Y$_p$ relays on a good determination of heavier
elements abundance, lets review the method commonly used for that purpose.

\begin{figure}[h]
\centering
\includegraphics
[height=9cm,width=9.5cm,angle=-90]{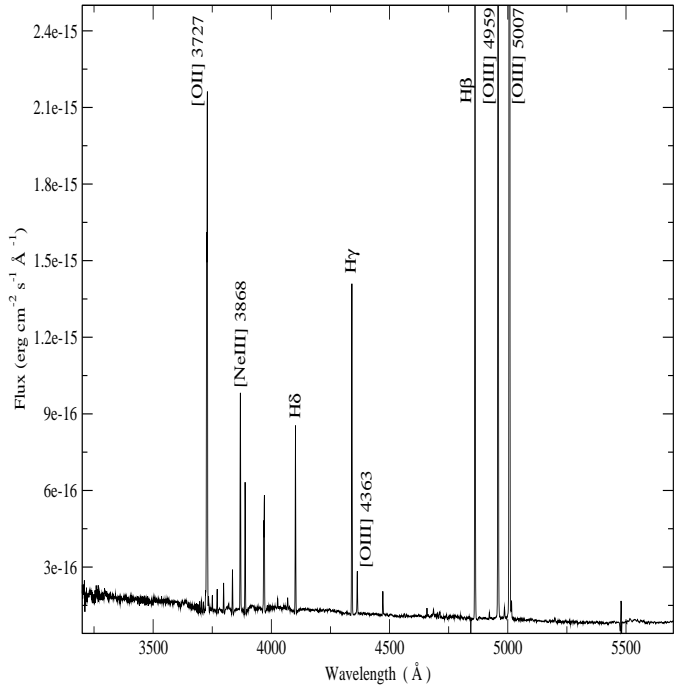}
\caption[h]{Optical blue 
spectrum of an  HII galaxy. The intense recombination and 
collisional forbidden lines are indicated. G.~H\"agele, private communication.}
\label{fig4}
\end{figure}

Chemical abundances in ionized regions of the interstellar medium (HII regions) 
are determined from their optical spectra, (an example of which is represented in 
Figure \ref{fig4}) applying relations of the type of equation 
\ref{eq.1}
to the emission-line spectrum as the intensity of a line is linked to
the ionic abundance through a function of the electron temperature (T$_e$).

\begin{equation}
      \label{eq.1}
\frac{N(X)}{N(H^+)}= f(T_e)\frac{I(\lambda)}{I(H\beta)}
\end{equation}

T$_e$ is found by the analysis of suitable temperature sensitive 
spectral features, e.g.~the 
[OIII]$\lambda\lambda$~4363/(4959,5007) ratio as it is easy to understand from atomic 
physics first principles.
The ionic abundances are then summed to 
obtain the chemical abundance of a particular element. For example,

\begin{equation}
      \label{eq.2}
\frac{N(He^+)}{N(H^+)} \simeq  T^{\alpha_1}\frac{I(\lambda 5876)}{I(H\beta)}
\end{equation}
\begin{equation}
      \label{eq.3}
\frac{N(He^{++})}{N(H^+)} \simeq  T^{\alpha_2}\frac{I(\lambda 4686)}{I(H\beta)}
\end{equation}
\begin{equation}
      \label{eq.4}
\frac{N(He)}{N(H)}=  \frac{N(He^+)}{N(H^+)} + \frac{N(He^{++})}{N(H^+)}
\end{equation}
gives the total abundance of He.

\subsection{Photoionization models}

\begin{figure}[h]
\centering
\includegraphics
[height=9cm,width=9.5cm]{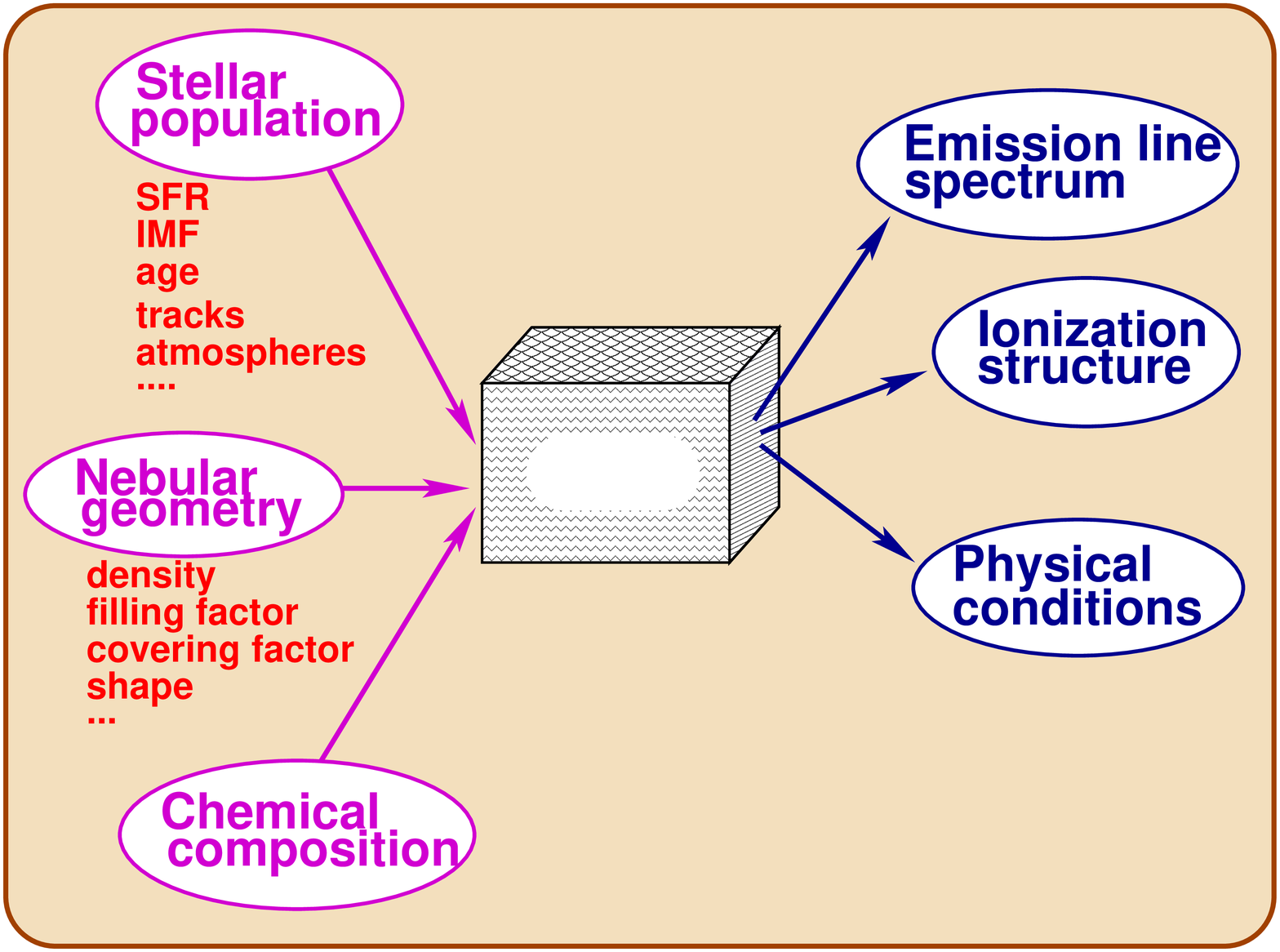}
\caption[h]{Schematic view of the way in which photoionization models
help us to deduce physical conditions of nebular gas.}
\label{fig5}
\end{figure}

In cases where the T$_e$ cannot be rigourously determined, one can resort
to more empirical methods, in particular, to photoionization models.
A schematic view of how they work is presented in figure \ref{fig5}.

The inputs to a photoionization code are
an ionizing spectrum, a nebular geometry and a chemical composition.
The outputs are an emission line spectrum, and the structure of the
nebula (ionization structure, electron density and temperature N$_e$, T$_e$). 
If we succeed in reproducing
the emission line spectrum of an observed object, we can use the
predicted structure to estimate properties such as the ionization
structure.

\section{Uncertainties in Y$_p$ determinations}

The uncertainties affecting Y$_p$ can be identified in terms 
of the individual steps that the method includes and  can be grouped into 
three broad categories: the physics, the stellar parameters and the nebular parameters.
We can
list some of them, but several others exist (e.g., observational errors).
In table \ref{t1} they are listed in three categories, according to 
the ingredient of the method where they originate. As it can be seen, some of them are 
intertwinned.
We will discuss all of them, except the physics that affect our abundance 
determinations through atomic parameters. There is nothing we astrophysicists can 
do about it, but to wait for better laboratory results.


\begin{table}[h]
\caption{Sources of uncertainty} 
\label{t1}
\begin{center}
\small
\begin{tabular}{c|l}

 Category & Source of error\\
\hline
& \\
physics &atomic parameters \\
\hline
& underlying  stellar absorption \\
stellar&  \\
parameters& --------------------------------------\\
---------------------------& ionization structure\\
nebular& \\
&temperature structure \\
parameters& \\
& H I collisional enhancement\\
\hline
\end{tabular}
\end{center}
\end{table}

\subsection{Underlying absorption}

The stars that photoionized the nebulae also show their own spectra. They are young 
hot stars, with H and He lines  in absorption, therefore the nebular emission 
measured is underestimated.
One solution is to subtract from the spectra a good estimate (obtained from 
theoretical stellar models) of the
stellar population. One is then  left with the pure emission spectrum.
Nowadays, state-of-the-art stellar population synthesis models with high spectral 
resolution allow us to perform a good underlying absorption correction.

\subsection{Ionization structure}
 
The simplest assumption to derive the He abundance from the HeI and HeII
abundances is that the ionized He and the ionized H spheres
coincide. If this is not the case, and the effect is not taken into account, 
the He abundance is either underestimated or overestimated 
depending on whether neutral helium is contained in the ionized hydrogen 
sphere, or H$^0$ is contained within the He$^0$ sphere.
If H{\sc II} regions were density-bounded in all 
directions, the problem would not exist.

There are several ways to deal with the uncertainty associated with the ionization 
structure:

\begin{itemize}
\item 1. applying selection criteria to the object sample.
\item 2. building tailored photoionization models.
\item 3. using narrow long-slit data.
\end{itemize}

A first possibility is applying selection criteria to the
regions analyzed. Since the ionization structure depends
on the shape of the ionizing spectrum, the selection criteria depend
on the stellar population. One can therefore exclude those
regions ionized by stellar clusters that do not guarantee  ionization correction 
factors ICF(He)$\sim$1. This has the disadvantage that precious  data points are lost, 
methods are not so robust as they should be,
and selection criteria are VERY model-dependent.

A second possibility is building tailored ionization models and
assuming that the ionization structure of the model reflects the one
of the real region. Disadvantages of this method are that it is 
time-consuming, and that the model must really be very constrained
to ensure that the predicted ICF is a good assessment.

A third possibility is using long-slit data. In this way, although the
problem is not solved, its impact is lessened owed to geometrical reasons
(the relative volume of the transition zone is smaller than in the case 
of a complete sphere, about  2/3).

\subsection{Temperature fluctuations}

Temperature fluctuations 
inside the H II regions can bias the abundance 
determinations. We use equation \ref{eq.1} to link the abundance of an element 
with the relative intensities of its lines to a hydrogen recombination line,
and the relation depends on a function of the electron temperature. But, does 
one value of T$_e$ fit all? The answer is no, as  
each ion is associated to a typical 
temperature, and adopting  one for all
introduces a bias in the derived abundance.

The dependence of the line intensity on T$_e$ varies from recombination to 
collisionally excited lines.
Recombination lines (like, e.g.~HeI) weight smoothly the T structure according to
\begin{equation}
      \label{eq.5}
\frac{I(\lambda 5876)}{I(H\beta)}\simeq T^{-\alpha} \frac{N(He^+)}{N(H^+)} 
\end{equation}
while collisional lines (like, e.g.~[OIII]$\lambda$ 5007\AA ) are enhanced 
in T peaks as

\begin{equation}
      \label{eq.6}
\frac{I(\lambda 5007)}{I(H\beta)} \simeq T^{-\beta} \times e^{\Delta E/kT} \times   
\frac{N(O^{++})}{N(H^+)}
\end{equation}
Therefore, if there are temperature fluctuations in the region
and the temperature is determined from the ratio of collisional lines,
as it is often the case, the inferred temperature will be biased
towards the peaks. On the other hand, recombination lines may give  
an estimate closer to the
average values. The problem is that recombination lines are weaker than
collisional lines, so they are more difficult to observe.

In any case, this becomes a very hairy problem. The temperature used to find 
the ionic abundances must be determined 
with care, otherwise the abundances will be 
over or underestimated.
 
\subsection{Collisional enhancement of Balmer lines}

Another problem is caused by the collisional enhancement of Balmer lines.
The expression here 
\begin{equation}
      \label{eq.7}
\frac{N(He^+)}{N(H^+)} \simeq  T^{\alpha}\frac{I(\lambda )}{I(H\beta)}
\end{equation}
holds under the hypothesis that the intensities
derive from recombination only. Helium intensities have an important
collisional part, which is routinely computed and subtracted out.
Balmer lines have a much smaller collisional part, which is nevertheless
important at the present required state of precission.
To have an idea of the relative importance of collisions in Balmer lines,
let us consider the collisional-to-recombination ratio of H$\beta$: the ratio
depends on the ionic fractions of neutral and ionized hydrogen and on the Boltzmann
factor of the transition from the ground state to the upper level of the line.
\begin{equation}
      \label{eq.8}
\frac{j(H\beta )_C}{j(H\beta )_R} \propto \frac{N(H^0)}{N(H^+)}e^{-\Delta E/KT}
\end{equation}
Typical values for the ionic ratio are around 10$^{-4}$,  values for the
Boltzmann factor are listed below for the typical range of HII temperatures.


\begin{table}[h]
\caption{Boltzman factor as a function of Temperature} 
\label{t2}
\begin{center}
\small
\begin{tabular}{c|l}

T$_e$& e$^{-\Delta E/KT}$\\
\hline
10000 & 3.7E-7\\
12500 & 7.2E-6 \\
15000 & 5.2E-5 \\
17500 & 2.1E-4 \\
20000 & 6.1E-4 \\
\hline
\end{tabular}
\end{center}
\end{table}

Due to the Boltzmann factor, the collisional contribution to the Balmer lines
is particularly important in high-temperature regions. This is a big
difficulty because the hottest regions are the most metal poor ones, 
that is the most important ones from the point of view of the determination
of primordial helium given that the amount of (unseen) He$^0$ is minimal, and so 
is the extrapolation to Z=0.

An additional difficulty is that the collisional contribution is higher
for H$\alpha$ than for H$\beta$, and thus it mimicks (and can be misunderstood for)
the reddening due the interstellar attenuation.
Therefore, unrecognized collisions have two main effects, one direct and one indirect.
The direct one is that the relative hydrogen contribution is overestimated,
therefore the He/H ratio is underestimated. The indirect one is that if the
Balmer decrement is interpreted in terms of pure extinction, the lines
blueward of H$\beta$ are overestimated, and those redward of H$\beta$
are underestimated. The global effect on helium abundance depends
on which lines are used, but since the most intense ones ($\lambda$ 5876 and 6678 \AA )
are redward of H$\beta$, the net effect is usually that He is underestimated.

As an example,
\cite{12} used tailored photoionization models to study quantitatively the effect of 
collisional excitation of Balmer lines on the determination of the
helium abundance (Y) in individual photoionized regions.  
The geometry of the model is very complex, simulating 
a filamentary structure with stellar sources
spread everywhere.The revised Y values for 
the five objects in their sample yield an increase of +0.0035 in Y$_p$, giving
Y$_p=0.2391\pm 0.0020$. 

Other recent results, \cite{13} use a numerous sample of good  signal-to-noise spectra,
in an effort to take  into account all the possible systematics. They obtain
Y$_p=0.2421 \pm 0.0021$.

\begin{figure}[h]
\centering
\includegraphics
[height=9cm,width=9.5cm]{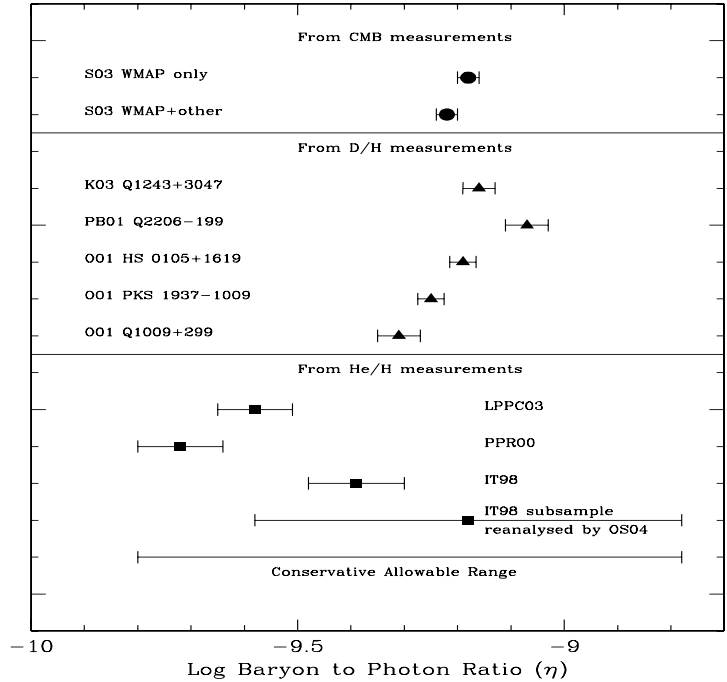}
\caption[h]{Concordance between WMAP derived light element abundances and 
observationally determined ones (from \cite{15} and references therein).}
\label{fig6}
\end{figure}

In a novel approach, \cite{14}
use the recent determinations of  the baryon density by the experiment WMAP 
(e.g.~\cite{15})
and the standard BBNS model to determine quite precise  predictions of the primordial
light-element abundances. They argue that the discrepancies between the 
observationally determined Y values and the value favoured by WMAP results are 
significant if only the statistical errors are considered.
They examine in detail some likely sources of systematic uncertainties that may 
resolve the differences between the determinations and conclude
``that the observational determination of the primordial helium abundance is 
completely limited by systematic errors and that these systematic errors have not been
fully accounted for in any published observational determination of the primordial 
helium abundance''. They advocate for a nonparametric approach to the analysis of
the observed metal-poor H II region spectra such that the physical conditions and the 
helium abundance are derived solely from the relative flux ratios of the
helium and hydrogen emission lines.
The data and the selected objects for the task should be very specific and of 
inexistent (at present) quality. In practice, the solutions at present are parametric
 with underestimated error bars. \cite{14} obtain a value of Y$_p=0.249\pm 0.009$. 
They consider as their main result  the increase in the size of the uncertainty rather 
than the shift in the primordial value and go on to claim that most of the spectra 
analyzed to date do not significantly constrain Y$_p$.
Rather, they argue that  a range of allowed
values would be  0.232$\leq Y_p\leq $0.258. 
Figure \ref{fig6}, taken from their work, shows that this last consideration
allows for the eagerly searched concordance between measurements of the 
baryon-to-photon ratio ($\eta$) from WMAP and  deuterium and helium abundances.
It also shows that individual error bars in D/H do not 
overlap. As a consequence, averaging the five best absorption
system determinations may not be correct.

\begin{figure}[h]
\centering
\includegraphics
[height=9.5cm,width=10.cm]{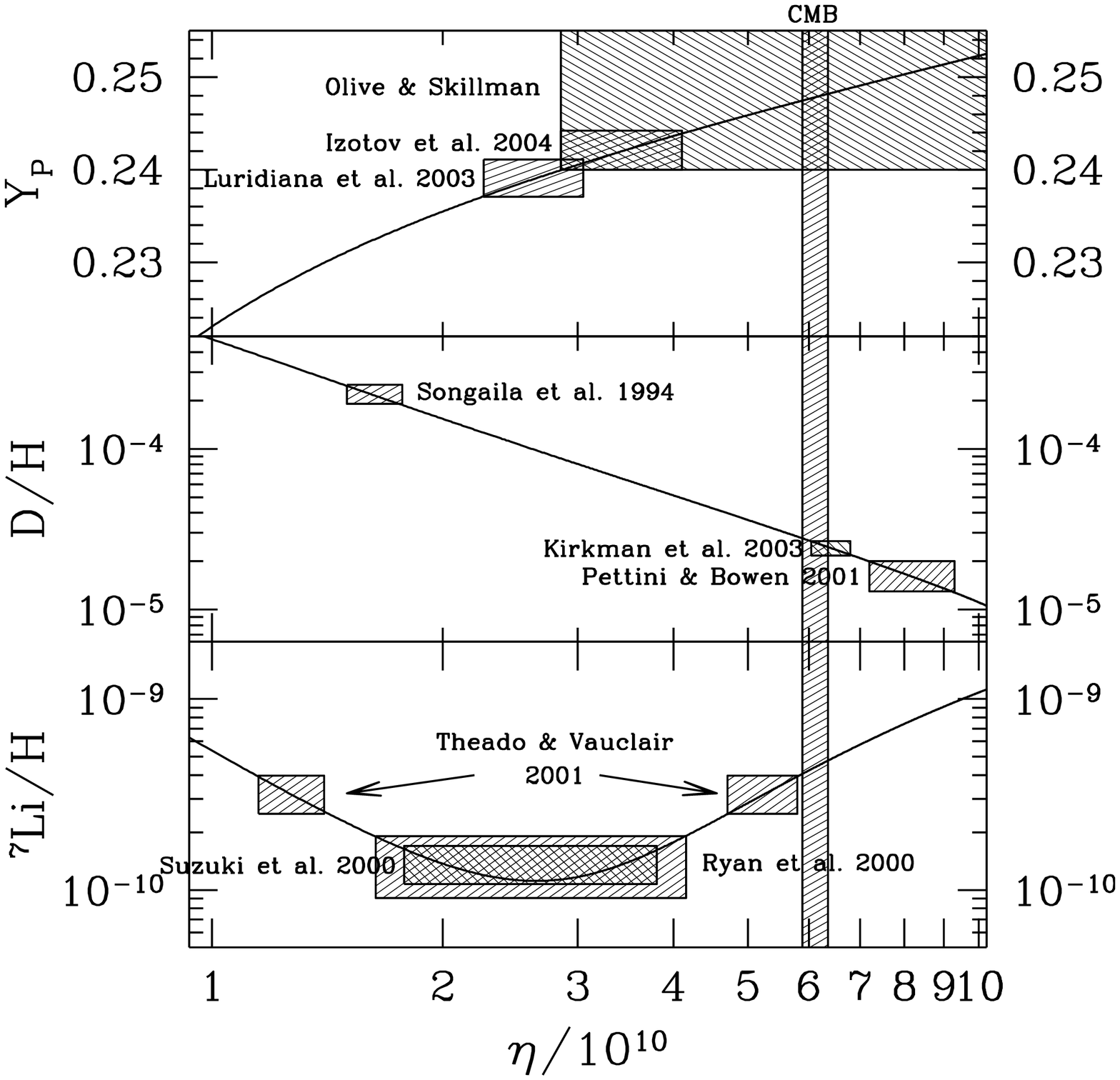}
\caption[h]{State of the art relation between the cosmological parameter $\eta$
and the primordial abundances of light elements.}
\label{fig7}
\end{figure}

Finally, figure  \ref{fig7} summarises the state-of-the-art concerning the determination
of $\eta$ vs.~the primordial abundances of the light elements. 

It is interesting to note that the  Songaila 1994 point for D/H
has been included just for historical reasons,
as it was soon recognised to be wrong.

\section{Concluding remarks}

Given the concordance obtained with WMAP ­ D/H 
determination (e.g.~\cite{15}),  is the $\eta$ value secure?
The answer is no.  Cosmic Microwave Background results constraint combinations of 
cosmological 
parameters, therefore independent individual determinations of them are still 
very important.

Concordance of primordial abundances has 
been considered a triumph of modern 
cosmology; regaining it represents an important goal.

In the process, as the comunity has done in the past, we are sure to learn a lot
about chemical evolution of galaxies and about systematic effects in abundance 
determinations.

\section*{Acknowledgments}
I am pleased to thank the organizers for their monumental efforts to make this 
Conference such a success. ET wants particularly to thank the opportunity  of
realizing the long-kept dream of getting to know the land of her ancestors. 
ET and RJT acknowledge the financial support given by the 
Mexican Research Council (CONACYT) through research grants 40018-A-1 
and E32186. 
Valentina Luridiana thanks the hospitality of INAOE 
during a visit in which  this work flourished.

\end{document}